\documentclass[aps,prl,bibnotes,reprint,twocolumn,showpacs,preprintnumbers,amsmath,amssymb,superscriptaddress,floatfix]{revtex4-2}
\usepackage[T1]{fontenc}
\usepackage[ansinew]{inputenc}
\usepackage{graphics}
\usepackage{dcolumn}
\usepackage{bm}
\usepackage{amsthm}
\usepackage{amsmath}
\usepackage{amssymb}
\usepackage{siunitx}
\usepackage{diagbox}
\usepackage{hyperref}
\usepackage{url}
\usepackage{xcolor}

\begin{document}

\title{Scaling in Magnetic Neutron Scattering}

\author{Venus Rai}
\author{Andreas Michels}\email[Electronic address: ]{andreas.michels@uni.lu}

\affiliation{Department of Physics and Materials Science, University of Luxembourg, 162A~avenue de la Faiencerie, 1511~Luxembourg, Grand Duchy of Luxembourg}


\begin{abstract}
We report the discovery of scaling in the mesoscale magnetic microstructure of bulk ferromagnets. Supported by analytical micromagnetic theory, we introduce the field-dependent scaling length $l_{\mathrm{C}}(H)$, which describes the characteristic long-wavelength magnetization fluctuations that are caused by microstructural defects by means of magnetoelastic and magnetocrystalline anisotropy. The scaling length $l_{\mathrm{C}}$ is identified to consist of the micromagnetic exchange length of the field $l_{\mathrm{H}}$, which depends on the magnetic interactions, and a field-independent contribution that reflects the properties of the magnetic anisotropy field and the magnetostatic fluctuations. The latter finding is rooted in the convolution relationship between the grain microstructure and micromagnetic response functions. We validated the scaling property by analyzing experimental data for the magnetic neutron scattering cross section. When plotted as a function of the dimensionless scaled scattering vector $\mathfrak{q}(H) = q \, l_{\mathrm{C}}(H)$, the field-dependent amplitude-scaled neutron data of nanocrystalline Co and a Nd-Fe-B-based nanocomposite collapse onto a single master curve, demonstrating universal behavior. The scaling length $l_{\mathrm{C}}$ provides a framework for analyzing the field-dependent neutron scattering cross section, highlighting the existence of critical length scales that govern the mesoscale microstructure of magnetic materials.
\end{abstract}

\date{\today}

\maketitle


{\it Introduction.} Scaling is a fundamental concept in physics (and in the natural sciences in general) that reveals how physical laws and phenomena change with size, time, energy, or other relevant variables~\cite{scalingbook1,scalingbook2}. It may serve as a bridge between different regimes of a system, offering deep insights into the underlying principles that govern diverse physical processes. Well-known examples from condensed-matter physics are second-order phase transitions, where scaling laws encapsulate the universal behavior of physical quantities such as the magnetic susceptibility or the correlation length~\cite{hohenberg1977,yeomans}, or the dynamical scaling of the phase separation process in binary metal alloys~\cite{dynscaleneutrons}. Other more recent examples for the relevance of scaling include the finding of a scaling law for the intrinsic fracture energy of stretchable networks~\cite{scaling_paper1}, the power-law scaling in neuronal networks on the example of the fruit fly brain~\cite{scaling_paper2}, a scaling law for how the timescale of solidification under homogeneous nucleation depends upon the compression rate in both metallic and molecular systems~\cite{scaling_paper3}, or the Kibble-Zurek scaling of the defect density as a function of the quench time (and deviations thereof) when crossing a continuous phase transition~\cite{scaling_paper4}.

\begin{figure}[tb!]
\centering
\resizebox{1.0\columnwidth}{!}{\includegraphics{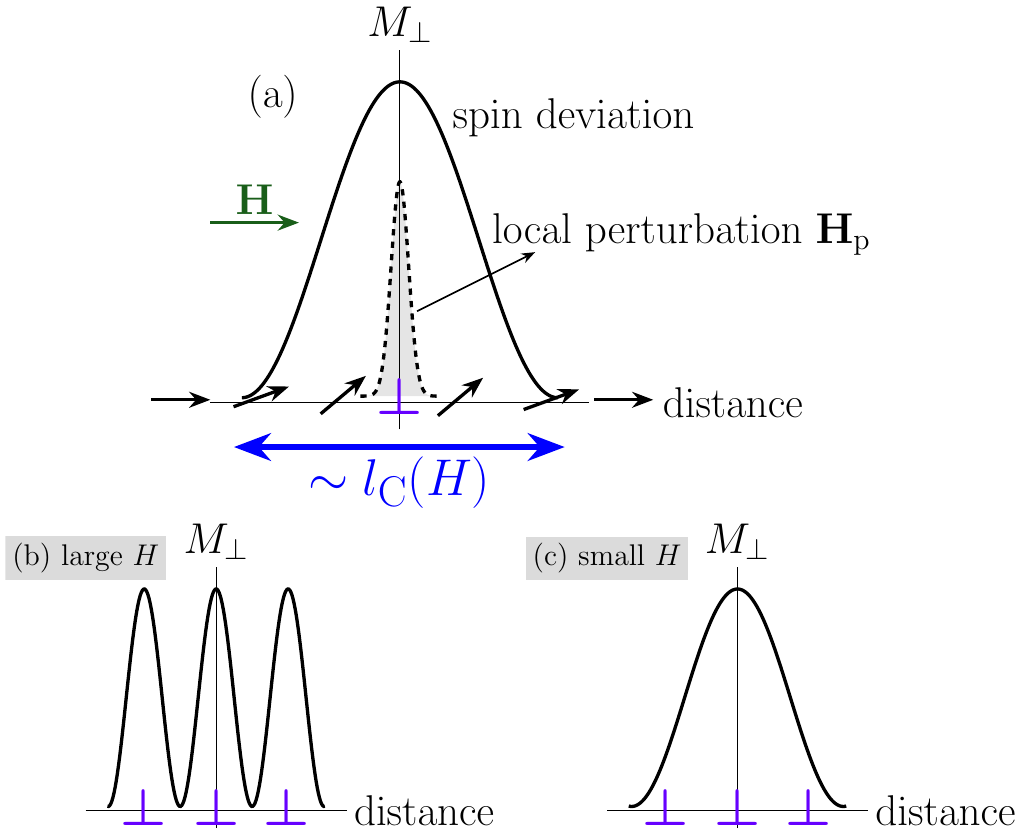}}
\caption{Illustration of the magnetic scaling concept. (a)~The length scale $l_{\mathrm{C}}(H)$ is a measure for the size of inhomogeneously magnetized regions around microstructural defects ($\perp$) (after~\cite{kronfahn03}). The latter are characterized by a magnetic anisotropy field $\mathbf{H}_{\mathrm{p}}(\mathbf{r})$ that is at the origin of the spin perturbation, e.g., due to magnetocrystalline or magnetoelastic anisotropy. $M_{\perp}$ denotes the component of the local magnetization vector $\mathbf{M}(\mathbf{r})$ that is perpendicular to the applied magnetic field $\mathbf{H}$. The ferromagnetic exchange interaction transmits the perturbation from the defect core into the surrounding crystal lattice. At a given $H$, $l_{\mathrm{C}}$ may be seen as the resolution limit of $\mathbf{M}(\mathbf{r})$. Panels~(b) and (c) (not to scale) schematically display the perpendicular magnetization distribution around defects in the high-field, small-amplitude limit~(b), when $l_{\mathrm{C}}$ is small, and in the low-field, large-amplitude case~(c) (large $l_{\mathrm{C}}$). In~(b), the magnetization can follow the local direction of the anisotropy field characterizing the defect, whereas in~(c) the defect group appears as a single superdefect; in other words, when decreasing the field [(b)$\rightarrow$(c)], the amplitude ($M_{\perp}$) and the characteristic wavelength ($l_{\mathrm{C}}$) of the magnetization ``ripple'' increases.}
\label{fig1}
\end{figure}

Here, we report the discovery of a new scaling law for the mesoscale magnetic microstructure of bulk ferromagnets. Specifically, we show both theoretically and experimentally that the perturbing effect of microstructural defects on the surrounding spin structure can be described by a unique field-dependent scaling variable $l_{\mathrm{C}}(H)$. This length scale, which naturally emerges in the micromagnetic continuum description of the magnetic microstructure, characterizes the size (wavelength) of nonuniformly magnetized regions around defects (compare Fig.~\ref{fig1}). It has its origin in the complicated convolution relationship between the defect microstructure and the magnetization distribution.

To experimentally demonstrate our finding, we use the scaling variable to describe the field-dependent magnetic neutron scattering structure factor of several magnetic materials. Small-angle neutron scattering (SANS) emerges here as a key experimental technique, offering insights into the magnetic microstructure within a range of $\sim$$1$$-$$1000 \, \mathrm{nm}$~\cite{michelsbook}. When the momentum transfer or scattering vector $q$ is scaled by $l_{\mathrm{C}}(H)$, according to
\begin{equation}
\label{qscalingdef1}
\mathfrak{q}(H) = q \, l_{\mathrm{C}}(H) ,
\end{equation}
the SANS cross sections measured at different applied magnetic fields collapse onto a single master curve.

We start the discussion by recalling the basic ideas behind the theory of magnetic SANS. The theoretical concepts will then be benchmarked by comparison to experimental neutron data on nanocrystalline Co and a Nd-Fe-B-based nanocomposite.

{\it Micromagnetic SANS theory:} The theory of magnetic SANS (see, e.g., Refs.~\cite{michels2013,michelsPRB2016,metlov2022}) is based on the following expression for the magnetic Gibbs free energy~\cite{brown}:
\begin{eqnarray}
\label{eq:classicalgibbs}
G = \iiint_V dV \left( \frac{A}{M_0^2} \left[ (\nabla M_x)^2 + (\nabla M_y)^2 + (\nabla M_z)^2 \right] \right. \nonumber \\ \left. 
+ \omega_{\mathrm{a}} - \mu_0 \mathbf{M} \cdot \mathbf{H} - \frac{1}{2} \mu_0 \mathbf{M} \cdot \mathbf{H}_{\mathrm{d}} \right) ,
\end{eqnarray}
where $\mathbf{M}(\mathbf{r}) = \{ M_x(\mathbf{r}), M_y(\mathbf{r}), M_z(\mathbf{r}) \}$ is the magnetization vector field with $M_0 = |\mathbf{M}|$ being the saturation magnetization. The first term in Eq.~(\ref{eq:classicalgibbs}) describes the stiffness of the spin system due to symmetric exchange with exchange constant $A$, $\nabla = \{ \partial/\partial x, \partial/\partial y, \partial/\partial z \}$ is the del operator, the second term $\omega_{\mathrm{a}} = \omega_{\mathrm{a}}(\mathbf{M}(\mathbf{r}))$ denotes the magnetic anisotropy energy density, $\mathbf{H} = \{ 0, 0, H \}$ is the externally applied magnetic field (assumed to be constant here), and $\mathbf{H}_{\mathrm{d}} = \mathbf{H}_{\mathrm{d}}(\mathbf{r}, \mathbf{M}(\mathbf{r}))$ represents the magnetostatic field created by the magnetization distribution ($\mu_0 = 4\pi 10^{-7} \, \mathrm{Tm/A}$).

As detailed e.g.\ in Refs.~\cite{michels2013,michelsPRB2016}, the linearization of the Euler-Lagrange differential equations that result from the variation of the functional Eq.~(\ref{eq:classicalgibbs}) yields the following closed-form expressions for the transversal magnetization Fourier components $\widetilde{M}_x(\mathbf{q})$ and $\widetilde{M}_y(\mathbf{q})$:
\begin{eqnarray}
\label{solmxgeneral}
\widetilde{M}_x &=& \frac{p \left( \widetilde{H}_{\mathrm{p}x} \left[ 1 + p \hat{q}_y^2 \right] - \widetilde{M}_z \hat{q}_x \hat{q}_z - \widetilde{H}_{\mathrm{p}y} p \hat{q}_x \hat{q}_y \right)}{1 + p \left( \hat{q}_x^2 + \hat{q}_y^2 \right)} , 
\end{eqnarray}
\begin{eqnarray}
\label{solmygeneral}
\widetilde{M}_y &=& \frac{p \left( \widetilde{H}_{\mathrm{p}y} \left[ 1 + p \hat{q}_x^2 \right] - \widetilde{M}_z \hat{q}_y \hat{q}_z - \widetilde{H}_{\mathrm{p}x} p \hat{q}_x \hat{q}_y \right)}{1 + p \left( \hat{q}_x^2 + \hat{q}_y^2 \right)} ,
\end{eqnarray}
where $\mathbf{\hat{q}} = \{ q_x, q_y, q_z \} / q$ denotes the unit wave vector (later on identified as the momentum-transfer vector),
\begin{equation}
\label{pdef}
p(q, H) = \frac{M_0}{H_{\rm eff}(q, H)}
\end{equation}
is a dimensionless function of $q = |\mathbf{q}|$ and $H = |\mathbf{H}|$, and
\begin{equation}
\label{heffdef}
H_{\mathrm{eff}}(q, H) = H ( 1 + l_{\mathrm{H}}^2 q^2 ) = H + \frac{2A}{\mu_0 M_0} q^2
\end{equation}
is the effective magnetic field, which contains the micromagnetic exchange length of the field
\begin{equation}
\label{lhdef}
l_{\mathrm{H}}(H) = \sqrt{\frac{2 A}{\mu_0 M_0 H}} .
\end{equation}
As we will see below, the field-dependent length scale $l_{\mathrm{H}}(H)$ forms an integral part of the scaling length $l_{\mathrm{C}}(H)$ introduced by Eq.~(\ref{qscalingdef1}). $\widetilde{H}_{\mathrm{p}x}(\mathbf{q})$ and $\widetilde{H}_{\mathrm{p}y}(\mathbf{q})$ represent the Cartesian Fourier components of the magnetic anisotropy field $\mathbf{H}_{\mathrm{p}}(\mathbf{r})$; these terms increase the magnitudes of $\widetilde{M}_{x,y}$ and tend to produce spin disorder in the system. Likewise, $\widetilde{M}_z(\mathbf{q})$ is the Fourier transform of the spatial saturation magnetization profile $M_{\mathrm{s}}(\mathbf{r})$ of the sample. Note that the volume average of $M_{\mathrm{s}}(\mathbf{r})$ equals the macroscopic saturation magnetization $M_0 = \langle M_{\mathrm{s}}(\mathbf{r}) \rangle$, which can be measured with a magnetometer. Note also the symmetry of the Eqs.~(\ref{solmxgeneral}) and (\ref{solmygeneral}) under the exchange of the $x$ and $y$~coordinates.

The effective magnetic field $H_{\mathrm{eff}}$ [Eq.~(\ref{heffdef})] consists of a contribution due to the applied field $H$ and of the exchange field $2 A q^2/ (\mu_0 M_0)$. An increase of $H$ increases $H_{\mathrm{eff}}$ only at the smallest $q$~values, whereas $H_{\mathrm{eff}}$ at the larger $q$ is always very large ($\sim$$10$$-$$100 \, \mathrm{T}$) and independent of $H$. The latter statement may be seen as a manifestation of the fact that exchange forces tend to dominate on small length scales~\cite{aharonibook}. Since $H_{\mathrm{eff}}$ appears predominantly in the denominators of the expressions for $\widetilde{M}_x$ and $\widetilde{M}_y$ [Eqs.~(\ref{solmxgeneral}) and (\ref{solmygeneral})], its role is to suppress the high-$q$ Fourier components of the magnetization, which correspond to sharp fluctuations in real space. However, long-range magnetization fluctuations, at small $q$, are effectively suppressed when $H$ is increased (compare the experimental magnetic SANS data in Fig.~\ref{fig2} below).

With a view towards the analysis of experimental neutron data, where most often the applied magnetic field $\mathbf{H}$ is perpendicular to the incoming neutron beam, we focus on the following purely magnetic SANS cross section~\cite{michelsbook}:
\begin{eqnarray}
\label{sigmasansperp2d}
\frac{d \Sigma_{\mathrm{M}}}{d \Omega} = \frac{8 \pi^3}{V} b_{\mathrm{H}}^2 \left( |\widetilde{M}_x|^2 + |\widetilde{M}_y|^2 \cos^2\theta \right. \\ \left. - (\widetilde{M}_y \widetilde{M}_z^{\ast} + \widetilde{M}_y^{\ast} \widetilde{M}_z) \sin\theta \cos\theta \right) \nonumber ,
\end{eqnarray}
where $V$ is the scattering volume, $b_{\mathrm{H}} = 2.91 \times 10^{8} \, \mathrm{A^{-1} m^{-1}}$ represents the atomic magnetic scattering length in small-angle approximation, $\mathbf{\widetilde{M}}(\mathbf{q}) = \{ \widetilde{M}_x(\mathbf{q}), \widetilde{M}_y(\mathbf{q}), \widetilde{M}_z(\mathbf{q}) \}$ is the Fourier transform of $\mathbf{M}(\mathbf{r})$, ``$\, ^{\ast} \,$'' refers to the complex conjugated quantity, and $\mathbf{\hat{q}} = \mathbf{q}/q \cong \{ 0, \sin\theta, \cos\theta\}$ is the unit scattering vector with $\theta$ the angle included between $\mathbf{H}$ and $\mathbf{\hat{q}}$. As shown in Ref.~\cite{michels2013}, near magnetic saturation, the azimuthally-averaged (over the detector plane) magnetic SANS cross section $d \Sigma_{\mathrm{M}} / d \Omega$ can be expressed in compact form as~\footnote{In the approach-to-saturation regime, $d\Sigma_{\mathrm{M}} / d \Omega$ equals the total (nuclear and magnetic) unpolarized SANS cross section minus the total SANS cross section at complete magnetic saturation.}:
\begin{equation}
\label{sigmasmperp}
\frac{d \Sigma_{\mathrm{M}}}{d \Omega}(q, H) = S_{\mathrm{H}}(q) \, R_{\mathrm{H}}(q, H) + S_{\mathrm{M}}(q) \, R_{\mathrm{M}}(q, H) ,
\end{equation}
where
\begin{equation}
\label{shdef}
S_{\mathrm{H}}= \frac{8 \pi^3}{V} b_{\mathrm{H}}^2 |\widetilde{H}_{\mathrm{p}}|^2 \,\,\, \mathrm{and} \,\,\, S_{\mathrm{M}} = \frac{8 \pi^3}{V} b_{\mathrm{H}}^2 |\widetilde{M}_z|^2
\end{equation}
denote, respectively, the anisotropy-field and magnetostatic scattering functions. $S_{\mathrm{H}}(q)$ and $S_{\mathrm{M}}(q)$---both field-independent in the approach-to-saturation regime---contain information on the strength and spatial structure of the magnetic anisotropy field and magnetostatic field; e.g., in a magnetic nanocomposite, $S_{\mathrm{M}} \propto |\widetilde{M}_z|^2 \propto (\Delta M)^2$, where $\Delta M$ denotes the jump of the magnetization magnitude at internal particle-matrix interfaces.


The dimensionless so-called micromagnetic response functions $R_{\mathrm{H}}(q, H)$ and $R_{\mathrm{M}}(q, H)$ are expressed as~\cite{michels2013}:
\begin{equation}
\label{rhdefperpradav}
R_{\mathrm{H}} = \frac{p^2}{4} \left( 2 + \frac{1}{\sqrt{1 + p}} \right) \,\,\, \mathrm{and} \,\,\, R_{\mathrm{M}} = \frac{\sqrt{1 + p} - 1}{2} .
\end{equation}
The magnetic neutron scattering due to transversal spin components, with related Fourier amplitudes $\widetilde{M}_x(\mathbf{q})$ and $\widetilde{M}_y(\mathbf{q})$, is contained in $d \Sigma_{\mathrm{M}} / d \Omega$, which decomposes into a term $S_{\mathrm{H}} R_{\mathrm{H}}$ due to perturbing magnetic anisotropy fields and a part $S_{\mathrm{M}} R_{\mathrm{M}}$ related to magnetostatic fields. 

{\it Analysis of experimental data.} The quantity $d \Sigma_{\mathrm{M}} / d \Omega$ [Eq.~(\ref{sigmasmperp})] represents, in the saturation regime, the $q$ and $H$~dependent unpolarized magnetic SANS cross section of a statistically-isotropic polycrystalline ferromagnet. In the following, we will show that, when the scattering vector $q$ is scaled by a suitably chosen length scale $l_{\mathrm{C}}$ [Eq.~(\ref{qscalingdef1})], the $d \Sigma_{\mathrm{M}} / d \Omega$ that are measured at a series of fields collapse onto a single master curve described by $d \Sigma_{\mathrm{M}} / d \Omega = \widetilde{\Sigma}_{\text{M}}(l_{\mathrm{C}}) d \Sigma_{\mathrm{M}} / d \Omega(\mathfrak{q})$, where the dimensionless vertical (amplitude) scaling is described by the quantity $\widetilde{\Sigma}_{\text{M}}$. For use in $\mathfrak{q} = q l_{\mathrm{C}}$, we make the following ansatz for the field-dependent correlation length:
\begin{equation}
\label{qscalingdef2}
l_{\mathrm{C}}(H) = f(\xi_{\mathrm{H}}, \xi_{\mathrm{M}}, \alpha) + l_{\mathrm{H}}(H) ,
\end{equation}
where the field-independent quantity $f$ characterizes the microstructure of the magnetic anisotropy (``H'') and magnetostatic (``M'') fields, and the field-dependent contribution $l_{\mathrm{H}}$ [Eq.~(\ref{lhdef})] depends on the magnetic interactions (exchange and magnetostatics). The parameters $\xi_{\mathrm{H}}, \xi_{\mathrm{M}}, \alpha$ determining $f$ are the correlation lengths of the magnetic anisotropy ($\xi_{\mathrm{H}}$) and magnetostatic ($\xi_{\mathrm{M}}$) fields and the relative strength $\alpha$ of the two contributions; for example, $\xi_{\mathrm{H}} = \xi_{\mathrm{M}}$ for a dilute collection of uniformly magnetized single-crystalline nanoparticles that are embedded in a homogeneous magnetic matrix of different magnetization.

\begin{figure*}[t!]
\centering
\resizebox{1.0\columnwidth}{!}{\includegraphics{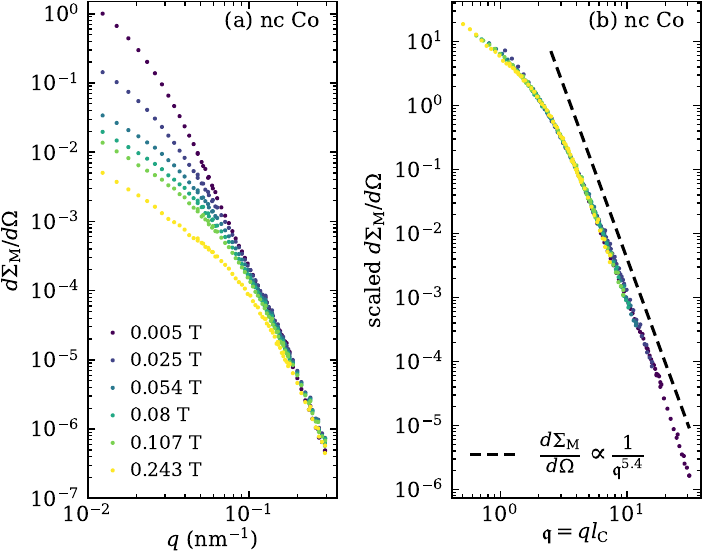}}
\resizebox{1.0\columnwidth}{!}{\includegraphics{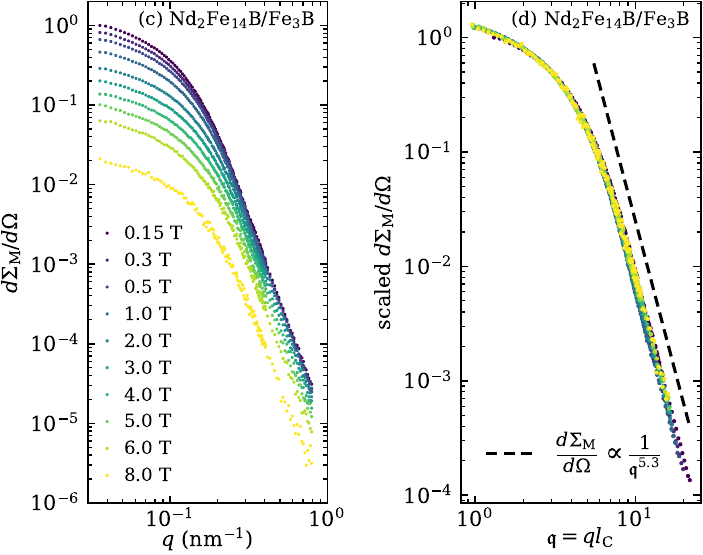}}
\caption{Scaling analysis of experimental magnetic neutron data. (a)~The field-dependent $\frac{d \Sigma_{\mathrm{M}}}{d \Omega}(q, H)$ of nanocrystalline Co (log-log scale, field values from top to bottom are specified in the inset) (data taken from Ref.~\cite{michels03prl}). (b)~Data from panel~(a) plotted as a function of $\mathfrak{q} = q l_{\mathrm{C}}$ with $f \cong 11.8 \pm 1.4 \, \mathrm{nm}$ and vertically scaled using the function $\widetilde{\Sigma}_{\mathrm{M}}(l_{\mathrm{C}})$ [compare Fig.~\ref{fig3}(b)]. (c)~$\frac{d \Sigma_{\mathrm{M}}}{d \Omega}(q, H)$ of the nanocomposite Nd$_{2}$Fe$_{14}$B/Fe$_{3}$B (log-log scale, field values from top to bottom are specified in the inset) (data taken from Ref.~\cite{bickapl2013}). (d)~Scaled magnetic SANS cross section $\frac{d \Sigma_{\mathrm{M}}}{d \Omega}(\mathfrak{q})$ using $f \cong 24.5 \pm 0.9 \, \mathrm{nm}$ and vertically scaled [Fig.~\ref{fig3}(b)]. The data in (a) and (c) were normalized by the intensity value at the smallest field and $q$~value, respectively, in this way making the $d \Sigma_{\mathrm{M}} / d \Omega$ data dimensionless. Dashed lines in (b) and (d):~asymptotic power laws $d \Sigma_{\mathrm{M}} / d \Omega \propto \mathfrak{q}^{-n}$ (see insets).}
\label{fig2}
\end{figure*}

The above ansatz for the scaling length [Eq.~(\ref{qscalingdef2})] can be physically motivated by inspecting Eqs.~(\ref{solmxgeneral}) and (\ref{solmygeneral}) for $\hat{q}_x = 0$ (corresponding to the scattering geometry in which the data in Fig.~\ref{fig2} were taken) and without dipolar interaction. In this situation we find~\cite{michelsbook}:
\begin{eqnarray}
\label{solmxgeneralqx0}
\widetilde{M}_x &=& p \widetilde{H}_{\mathrm{p}x} = \frac{M_0}{H} \frac{\widetilde{H}_{\mathrm{p}x}}{ 1 + l_{\mathrm{H}}^2 q^2 } , 
\end{eqnarray}
\begin{eqnarray}
\label{solmygeneralqx0}
\widetilde{M}_y &=& \frac{M_0}{H} \frac{\widetilde{H}_{\mathrm{p}y}}{ 1 + l_{\mathrm{H}}^2 q^2 } .
\end{eqnarray}
Equations~(\ref{solmxgeneralqx0}) and (\ref{solmygeneralqx0}) imply that the (perpendicular) magnetic microstructure in real space, $\mathbf{M}_{\perp}(\mathbf{r})$, corresponds (at not too small distances) to the convolution ($\ast$, not to be confused with the complex conjugated) of the anisotropy field microstructure, $\mathbf{H}_{\mathrm{p}}(\mathbf{r})$, with an exponential response function that decays with a characteristic length scale $l_{\mathrm{H}}$, i.e.,
\begin{eqnarray}
\label{convprop}
\mathbf{M}_{\perp}(\mathbf{r}) \cong \mathbf{H}_{\mathrm{p}}(\mathbf{r}) \ast \exp(-r/l_{\mathrm{H}}) .
\end{eqnarray}
This consideration then motivates the above choice for the scaling length $l_{\mathrm{C}}$ [Eq.~(\ref{qscalingdef2})] to consist of a field-dependent term $l_{\mathrm{H}} \propto 1/\sqrt{H}$, which takes the magnetic interactions ($A, M_0$) into account, and a field-independent contribution $f$, which describes the size, shape, and magnitude ($\alpha, \xi_{\mathrm{H}}, \xi_{\mathrm{M}}$) of the defect that causes the spin perturbation [compare also to Fig.~\ref{fig1}(a)]. Due to the complexity (nonlinearity) of the magnetic microstructure more complex relationships than the assumed linear dependency $l_{\mathrm{C}} \propto l_{\mathrm{H}}$ are of course feasible.


Figure~\ref{fig2} displays the results of a scaling analysis of experimental neutron data. For this we have used unpolarized magnetic SANS data of nanocrystalline Co~\cite{michels03prl} and a Nd-Fe-B alloy~\cite{bickapl2013}. For the computation of the exchange length $l_{\mathrm{H}}$ in $\mathfrak{q} = q (f + l_{\mathrm{H}})$, we have used the experimental values for the exchange constants and saturation magnetizations~\cite{michels03prl,bickapl2013}: $A = 31 \, \mathrm{pJ/m}$ (Co), $A = 12.5 \, \mathrm{pJ/m}$ (Nd$_{2}$Fe$_{14}$B/Fe$_{3}$B), $\mu_0 M_0 = 1.76 \, \mathrm{T}$ (Co), and $\mu_0 M_0 = 1.60 \, \mathrm{T}$ (Nd$_{2}$Fe$_{14}$B/Fe$_{3}$B); see also Refs.~\cite{jprb2001,bickapl2013no2} for further details on these samples (magnetization, x-ray diffraction, and electron microscopy). Therefore, the value of $l_{\mathrm{H}}$ is fixed for a given value of the field. For the function $f(\xi_{\mathrm{H}}, \xi_{\mathrm{M}}, \alpha)$ we have made the simplest possible choice, i.e., we set $f$ equal to a constant---the defect size. This quantity was determined by a least-squares analysis, i.e., for all the experimental $\frac{d \Sigma_{\mathrm{M}}}{d \Omega}(q, H)$ data in Figs.~\ref{fig2}(a) and \ref{fig2}(c) the parameter $f$ was refined in $\mathfrak{q} = q (f + l_{\mathrm{H}})$ to minimize the mean square deviation between all the data points (one free $f$~parameter per neutron data set). While the {\it horizontal} scaling is done via $f$, the {\it vertical} scaling is performed using the function $\widetilde{\Sigma}_{\text{M}}(l_{\mathrm{C}})$.

The (normalized) experimental magnetic SANS cross sections [Figs.~\ref{fig2}(a) and \ref{fig2}(c)] exhibit a strong field dependence, in particular at the smallest momentum transfers $q$. As discussed earlier, this is related to the fact that long-range magnetization fluctuations, at small $q$, are effectively suppressed when $H$ is increased. Closer inspection of the field-dependent data reveals the existence of a field-dependent length scale that decreases with increasing field. This becomes evident from the observation that the point with the largest curvature in $d \Sigma_{\mathrm{M}} / d \Omega$ evolves to a larger $q$ when $H$ increases. Scaling of the field-dependent SANS data results in a collapse of all data onto a single master curve [Figs.~\ref{fig2}(b) and \ref{fig2}(d)]. Moreover, we see in Figs.~\ref{fig2}(b) and \ref{fig2}(d) that the magnetic structure factors are described by asymptotic power laws $d \Sigma_{\mathrm{M}} / d \Omega \propto \mathfrak{q}^{-n}$ that are much larger than the Porod sharp-interface exponent of $n=4$. This is in perfect agreement with the notion of spin-misalignment scattering, encapsulated in the magnetic SANS theory~\footnote{Note that asymptotically $R_{\mathrm{H}} \propto q^{-4}$ and $R_{\mathrm{M}} \propto q^{-2}$ in Eq.~(\ref{sigmasmperp}), which together with the (unknown) $q$~dependencies of $S_{\mathrm{H}}$ and $S_{\mathrm{M}}$ result in $d \Sigma_{\mathrm{M}} / d \Omega \propto q^{-n}$ with $n > 4$.}.

\begin{figure}[tb!]
\centering
\resizebox{1.0\columnwidth}{!}{\includegraphics{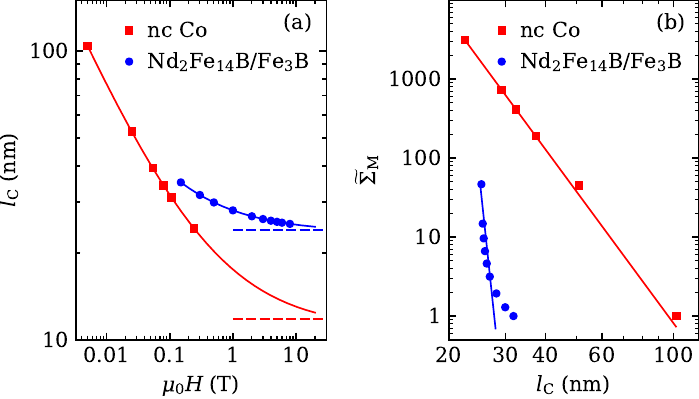}}
\caption{(a)~Field dependence of the scaling length $l_{\mathrm{C}}(H)$ for nanocrystalline Co and Nd$_{2}$Fe$_{14}$B/Fe$_{3}$B (log-log scale). Solid lines: $l_{\mathrm{C}}(H) = f + l_{\mathrm{H}}(H)$. The dashed lines indicate the respective defect size $l_{\mathrm{C}}(H \rightarrow \infty) = f$. (b)~Vertical scaling factor $\widetilde{\Sigma}_{\mathrm{M}}(l_{\mathrm{C}})$ and power-law fit (solid lines) to $\widetilde{\Sigma}_{\mathrm{M}}(l_{\mathrm{C}}) \propto l_{\mathrm{C}}^{-m}$ (semilog scale), where $m = 5.5 \pm 0.1$ for nanocrystalline Co.}
\label{fig3}
\end{figure}

The $l_{\mathrm{C}}(H)$~data in Fig.~\ref{fig3}(a) show a decrease of the characteristic spin-misalignment fluctuations with increasing field. The limiting values at large fields $l_{\mathrm{C}}(H \rightarrow \infty) = f$ (dashed lines) agree very well with previously determined structural features in the samples, i.e., an average grain size of $\sim$$10 \, \mathrm{nm}$ for nanocrystalline Co~\cite{jprb2001} and particle sizes for the two phases in Nd$_{2}$Fe$_{14}$B/Fe$_{3}$ between about $20$$-$$30 \, \mathrm{nm}$~\cite{bickapl2013no2}. These results suggest the interpretation of the limiting length scale $f$ as the average size of the defect that causes the spin disorder and the ensuing neutron scattering signal (compare to Fig.~\ref{fig1}). In experimental situations, the exchange constant $A$ may be determined from the field dependence of $l_{\mathrm{C}}$.

The vertical scale factor $\widetilde{\Sigma}_{\mathrm{M}}$ as a function of $l_{\mathrm{C}}$ is shown in Fig.~\ref{fig3}(b). For nanocrystalline Co, the amplitude can be well described by a power law $\widetilde{\Sigma}_{\mathrm{M}}(l_{\mathrm{C}}) \propto l_{\mathrm{C}}^{-m}$ with an exponent of $m = 5.5$. For Nd$_{2}$Fe$_{14}$B/Fe$_{3}$B, the agreement is less quantitative and no exponent can be reliably determined. This observation might be related to the fact that the small-misalignment approximation (that underlies our analytical SANS theory) becomes less reliable at small fields (large $l_{\mathrm{C}}$). Here, numerical micromagnetic computations, which are able to take into account the full nonlinearity of Brown's equations at low fields, might provide an extension of the mesoscale scaling concept in magnetic neutron scattering.  

Overall, the picture of the magnetic microstructure that emerges is the following: the defects (grains) are locally decorated by nanoscale spin disorder, which is generated by spatial variations in the direction and magnitude of the magnetic anisotropy field and by spatial variations in the magnetic materials parameters (e.g., exchange constant, saturation magnetization). At large fields, the scaling length $l_{\mathrm{C}}$ has a magnitude that is close to the defect size, while decreasing the field results in the build-up of long-wavelengths magnetization fluctuations. These are particularly large for the case of nanocrystalline Co ($l_{\mathrm{C}} \sim 100 \, \mathrm{nm}$ at $5 \, \mathrm{mT}$), suggesting that many grains in an exchange-coupled volume act as a single superdefect. The whole process is governed by a single field-dependent length scale $l_{\mathrm{C}}$.

{\it Conclusion.} Our analytical and experimental investigations of the magnetic neutron scattering cross section provide strong evidence for scaling behavior in the mesoscopic magnetic microstructure of bulk ferromagnets. This scaling arises from the convolution relationship between the grain microstructure and micromagnetic response functions, which govern the magnetization distribution. The characteristic scaling length consists of a field-independent contribution, reflecting the intrinsic properties of the defect responsible for spin perturbation, and a field-dependent micromagnetic exchange length that governs the propagation of the perturbation into the surrounding microstructure. We experimentally demonstrate the existence of this scaling in two distinct nanocrystalline magnetic systems:~a single-phase elemental ferromagnet and a two-phase nanocomposite. Our results establish a novel conceptual framework for analyzing field-dependent small-angle neutron scattering data, enabling a more precise interpretation of experimental results. Notably, the scaling length can be directly estimated from structural defect characteristics---accessible via electron microscopy or x-ray diffraction---and known magnetic material parameters. The defects arise due to the growth of the material and their density can be tuned e.g.\ by mechanical deformation or annealing. Furthermore, for previously uncharacterized materials, this approach provides a method to determine the exchange constant from the decay of the scaling length.

Critical reading of the manuscript by Hamid Kachkachi and Adolfo del Campo is gratefully acknowledged.


\begin{thebibliography}{22}%
\makeatletter
\providecommand \@ifxundefined [1]{%
 \@ifx{#1\undefined}
}%
\providecommand \@ifnum [1]{%
 \ifnum #1\expandafter \@firstoftwo
 \else \expandafter \@secondoftwo
 \fi
}%
\providecommand \@ifx [1]{%
 \ifx #1\expandafter \@firstoftwo
 \else \expandafter \@secondoftwo
 \fi
}%
\providecommand \natexlab [1]{#1}%
\providecommand \enquote  [1]{``#1''}%
\providecommand \bibnamefont  [1]{#1}%
\providecommand \bibfnamefont [1]{#1}%
\providecommand \citenamefont [1]{#1}%
\providecommand \href@noop [0]{\@secondoftwo}%
\providecommand \href [0]{\begingroup \@sanitize@url \@href}%
\providecommand \@href[1]{\@@startlink{#1}\@@href}%
\providecommand \@@href[1]{\endgroup#1\@@endlink}%
\providecommand \@sanitize@url [0]{\catcode `\\12\catcode `\$12\catcode `\&12\catcode `\#12\catcode `\^12\catcode `\_12\catcode `\%12\relax}%
\providecommand \@@startlink[1]{}%
\providecommand \@@endlink[0]{}%
\providecommand \url  [0]{\begingroup\@sanitize@url \@url }%
\providecommand \@url [1]{\endgroup\@href {#1}{\urlprefix }}%
\providecommand \urlprefix  [0]{URL }%
\providecommand \Eprint [0]{\href }%
\providecommand \doibase [0]{https://doi.org/}%
\providecommand \selectlanguage [0]{\@gobble}%
\providecommand \bibinfo  [0]{\@secondoftwo}%
\providecommand \bibfield  [0]{\@secondoftwo}%
\providecommand \translation [1]{[#1]}%
\providecommand \BibitemOpen [0]{}%
\providecommand \bibitemStop [0]{}%
\providecommand \bibitemNoStop [0]{.\EOS\space}%
\providecommand \EOS [0]{\spacefactor3000\relax}%
\providecommand \BibitemShut  [1]{\csname bibitem#1\endcsname}%
\let\auto@bib@innerbib\@empty
\bibitem [{\citenamefont {Fr\"ohlich~(editor)}(1983)}]{scalingbook1}%
  \BibitemOpen
  \bibfield  {author} {\bibinfo {author} {\bibfnamefont {J.}~\bibnamefont {Fr\"ohlich~(editor)}},\ }\href {https://doi.org/https://doi.org/10.1007/978-1-4899-6762-6} {\emph {\bibinfo {title} {{Scaling and Self-Similarity in Physics: Renormalization in Statistical Mechanics and Dynamics}}}}\ (\bibinfo  {publisher} {Springer},\ \bibinfo {address} {Basel},\ \bibinfo {year} {1983})\BibitemShut {NoStop}%
\bibitem [{\citenamefont {Barenblatt}(1996)}]{scalingbook2}%
  \BibitemOpen
  \bibfield  {author} {\bibinfo {author} {\bibfnamefont {G.~I.}\ \bibnamefont {Barenblatt}},\ }\href {https://doi.org/https://doi.org/10.1017/CBO9781107050242} {\emph {\bibinfo {title} {{Scaling, self-similarity, and intermediate asymptotics}}}}\ (\bibinfo  {publisher} {Cambridge University Press},\ \bibinfo {address} {Cambridge},\ \bibinfo {year} {1996})\BibitemShut {NoStop}%
\bibitem [{\citenamefont {Hohenberg}\ and\ \citenamefont {Halperin}(1977)}]{hohenberg1977}%
  \BibitemOpen
  \bibfield  {author} {\bibinfo {author} {\bibfnamefont {P.~C.}\ \bibnamefont {Hohenberg}}\ and\ \bibinfo {author} {\bibfnamefont {B.~I.}\ \bibnamefont {Halperin}},\ }\bibfield  {title} {\bibinfo {title} {{Theory of dynamic critical phenomena}},\ }\href {https://doi.org/10.1103/RevModPhys.49.435} {\bibfield  {journal} {\bibinfo  {journal} {Rev. Mod. Phys.}\ }\textbf {\bibinfo {volume} {49}},\ \bibinfo {pages} {435} (\bibinfo {year} {1977})}\BibitemShut {NoStop}%
\bibitem [{\citenamefont {Yeomans}(1992)}]{yeomans}%
  \BibitemOpen
  \bibfield  {author} {\bibinfo {author} {\bibfnamefont {J.~M.}\ \bibnamefont {Yeomans}},\ }\href@noop {} {\emph {\bibinfo {title} {{Statistical Mechanics of Phase Transitions}}}}\ (\bibinfo  {publisher} {Clarendon Press},\ \bibinfo {address} {Oxford},\ \bibinfo {year} {1992})\BibitemShut {NoStop}%
\bibitem [{\citenamefont {Katano}\ and\ \citenamefont {Iizumi}(1984)}]{dynscaleneutrons}%
  \BibitemOpen
  \bibfield  {author} {\bibinfo {author} {\bibfnamefont {S.}~\bibnamefont {Katano}}\ and\ \bibinfo {author} {\bibfnamefont {M.}~\bibnamefont {Iizumi}},\ }\bibfield  {title} {\bibinfo {title} {{Crossover Phenomenon in Dynamical Scaling of Phase Separation in Fe-Cr Alloy}},\ }\href {https://doi.org/10.1103/PhysRevLett.52.835} {\bibfield  {journal} {\bibinfo  {journal} {Phys. Rev. Lett.}\ }\textbf {\bibinfo {volume} {52}},\ \bibinfo {pages} {835} (\bibinfo {year} {1984})}\BibitemShut {NoStop}%
\bibitem [{\citenamefont {Hartquist}\ \emph {et~al.}(2025)\citenamefont {Hartquist}, \citenamefont {Wang}, \citenamefont {Cui}, \citenamefont {Matusik}, \citenamefont {Deng},\ and\ \citenamefont {Zhao}}]{scaling_paper1}%
  \BibitemOpen
  \bibfield  {author} {\bibinfo {author} {\bibfnamefont {C.}~\bibnamefont {Hartquist}}, \bibinfo {author} {\bibfnamefont {S.}~\bibnamefont {Wang}}, \bibinfo {author} {\bibfnamefont {Q.}~\bibnamefont {Cui}}, \bibinfo {author} {\bibfnamefont {W.}~\bibnamefont {Matusik}}, \bibinfo {author} {\bibfnamefont {B.}~\bibnamefont {Deng}},\ and\ \bibinfo {author} {\bibfnamefont {X.}~\bibnamefont {Zhao}},\ }\bibfield  {title} {\bibinfo {title} {{Scaling Law for Intrinsic Fracture Energy of Diverse Stretchable Networks}},\ }\href {https://doi.org/10.1103/PhysRevX.15.011002} {\bibfield  {journal} {\bibinfo  {journal} {Phys. Rev. X}\ }\textbf {\bibinfo {volume} {15}},\ \bibinfo {pages} {011002} (\bibinfo {year} {2025})}\BibitemShut {NoStop}%
\bibitem [{\citenamefont {Zhang}\ \emph {et~al.}(2024)\citenamefont {Zhang}, \citenamefont {Moore}, \citenamefont {Ru},\ and\ \citenamefont {Yan}}]{scaling_paper2}%
  \BibitemOpen
  \bibfield  {author} {\bibinfo {author} {\bibfnamefont {X.-Y.}\ \bibnamefont {Zhang}}, \bibinfo {author} {\bibfnamefont {J.~M.}\ \bibnamefont {Moore}}, \bibinfo {author} {\bibfnamefont {X.}~\bibnamefont {Ru}},\ and\ \bibinfo {author} {\bibfnamefont {G.}~\bibnamefont {Yan}},\ }\bibfield  {title} {\bibinfo {title} {{Geometric Scaling Law in Real Neuronal Networks}},\ }\href {https://doi.org/10.1103/PhysRevLett.133.138401} {\bibfield  {journal} {\bibinfo  {journal} {Phys. Rev. Lett.}\ }\textbf {\bibinfo {volume} {133}},\ \bibinfo {pages} {138401} (\bibinfo {year} {2024})}\BibitemShut {NoStop}%
\bibitem [{\citenamefont {Myint}\ \emph {et~al.}(2023)\citenamefont {Myint}, \citenamefont {Sterbentz}, \citenamefont {Brown}, \citenamefont {Stoltzfus}, \citenamefont {Delplanque},\ and\ \citenamefont {Belof}}]{scaling_paper3}%
  \BibitemOpen
  \bibfield  {author} {\bibinfo {author} {\bibfnamefont {P.~C.}\ \bibnamefont {Myint}}, \bibinfo {author} {\bibfnamefont {D.~M.}\ \bibnamefont {Sterbentz}}, \bibinfo {author} {\bibfnamefont {J.~L.}\ \bibnamefont {Brown}}, \bibinfo {author} {\bibfnamefont {B.~S.}\ \bibnamefont {Stoltzfus}}, \bibinfo {author} {\bibfnamefont {J.-P.~R.}\ \bibnamefont {Delplanque}},\ and\ \bibinfo {author} {\bibfnamefont {J.~L.}\ \bibnamefont {Belof}},\ }\bibfield  {title} {\bibinfo {title} {{Scaling Law for the Onset of Solidification at Extreme Undercooling}},\ }\href {https://doi.org/10.1103/PhysRevLett.131.106101} {\bibfield  {journal} {\bibinfo  {journal} {Phys. Rev. Lett.}\ }\textbf {\bibinfo {volume} {131}},\ \bibinfo {pages} {106101} (\bibinfo {year} {2023})}\BibitemShut {NoStop}%
\bibitem [{\citenamefont {Zeng}\ \emph {et~al.}(2023)\citenamefont {Zeng}, \citenamefont {Xia},\ and\ \citenamefont {del Campo}}]{scaling_paper4}%
  \BibitemOpen
  \bibfield  {author} {\bibinfo {author} {\bibfnamefont {H.-B.}\ \bibnamefont {Zeng}}, \bibinfo {author} {\bibfnamefont {C.-Y.}\ \bibnamefont {Xia}},\ and\ \bibinfo {author} {\bibfnamefont {A.}~\bibnamefont {del Campo}},\ }\bibfield  {title} {\bibinfo {title} {{Universal Breakdown of Kibble-Zurek Scaling in Fast Quenches across a Phase Transition}},\ }\href {https://doi.org/10.1103/PhysRevLett.130.060402} {\bibfield  {journal} {\bibinfo  {journal} {Phys. Rev. Lett.}\ }\textbf {\bibinfo {volume} {130}},\ \bibinfo {pages} {060402} (\bibinfo {year} {2023})}\BibitemShut {NoStop}%
\bibitem [{\citenamefont {Kronm\"uller}\ and\ \citenamefont {F\"ahnle}(2003)}]{kronfahn03}%
  \BibitemOpen
  \bibfield  {author} {\bibinfo {author} {\bibfnamefont {H.}~\bibnamefont {Kronm\"uller}}\ and\ \bibinfo {author} {\bibfnamefont {M.}~\bibnamefont {F\"ahnle}},\ }\href@noop {} {\emph {\bibinfo {title} {{Micromagnetism and the Microstructure of Ferromagnetic Solids}}}}\ (\bibinfo  {publisher} {Cambridge University Press},\ \bibinfo {address} {Cambridge},\ \bibinfo {year} {2003})\BibitemShut {NoStop}%
\bibitem [{\citenamefont {Michels}(2021)}]{michelsbook}%
  \BibitemOpen
  \bibfield  {author} {\bibinfo {author} {\bibfnamefont {A.}~\bibnamefont {Michels}},\ }\href@noop {} {\emph {\bibinfo {title} {{Magnetic Small-Angle Neutron Scattering: A Probe for Mesoscale Magnetism Analysis}}}}\ (\bibinfo  {publisher} {Oxford University Press},\ \bibinfo {address} {Oxford},\ \bibinfo {year} {2021})\BibitemShut {NoStop}%
\bibitem [{\citenamefont {Honecker}\ and\ \citenamefont {Michels}(2013)}]{michels2013}%
  \BibitemOpen
  \bibfield  {author} {\bibinfo {author} {\bibfnamefont {D.}~\bibnamefont {Honecker}}\ and\ \bibinfo {author} {\bibfnamefont {A.}~\bibnamefont {Michels}},\ }\bibfield  {title} {\bibinfo {title} {{Theory of magnetic small-angle neutron scattering of two-phase ferromagnets}},\ }\href {https://doi.org/10.1103/PhysRevB.87.224426} {\bibfield  {journal} {\bibinfo  {journal} {Phys. Rev. B}\ }\textbf {\bibinfo {volume} {87}},\ \bibinfo {pages} {224426} (\bibinfo {year} {2013})}\BibitemShut {NoStop}%
\bibitem [{\citenamefont {Michels}\ \emph {et~al.}(2016)\citenamefont {Michels}, \citenamefont {Mettus}, \citenamefont {Honecker},\ and\ \citenamefont {Metlov}}]{michelsPRB2016}%
  \BibitemOpen
  \bibfield  {author} {\bibinfo {author} {\bibfnamefont {A.}~\bibnamefont {Michels}}, \bibinfo {author} {\bibfnamefont {D.}~\bibnamefont {Mettus}}, \bibinfo {author} {\bibfnamefont {D.}~\bibnamefont {Honecker}},\ and\ \bibinfo {author} {\bibfnamefont {K.~L.}\ \bibnamefont {Metlov}},\ }\bibfield  {title} {\bibinfo {title} {{Effect of Dzyaloshinski-Moriya interaction on elastic small-angle neutron scattering}},\ }\href {https://doi.org/10.1103/PhysRevB.94.054424} {\bibfield  {journal} {\bibinfo  {journal} {Phys. Rev. B}\ }\textbf {\bibinfo {volume} {94}},\ \bibinfo {pages} {054424} (\bibinfo {year} {2016})}\BibitemShut {NoStop}%
\bibitem [{\citenamefont {Zaporozhets}\ \emph {et~al.}(2022)\citenamefont {Zaporozhets}, \citenamefont {Oba}, \citenamefont {Michels},\ and\ \citenamefont {Metlov}}]{metlov2022}%
  \BibitemOpen
  \bibfield  {author} {\bibinfo {author} {\bibfnamefont {V.~D.}\ \bibnamefont {Zaporozhets}}, \bibinfo {author} {\bibfnamefont {Y.}~\bibnamefont {Oba}}, \bibinfo {author} {\bibfnamefont {A.}~\bibnamefont {Michels}},\ and\ \bibinfo {author} {\bibfnamefont {K.~L.}\ \bibnamefont {Metlov}},\ }\bibfield  {title} {\bibinfo {title} {{Small-angle neutron scattering by spatially inhomogeneous ferromagnets with a nonzero average uniaxial anisotropy}},\ }\href {https://doi.org/10.1107/S160057672200437X} {\bibfield  {journal} {\bibinfo  {journal} {J. Appl. Cryst.}\ }\textbf {\bibinfo {volume} {55}},\ \bibinfo {pages} {592} (\bibinfo {year} {2022})}\BibitemShut {NoStop}%
\bibitem [{\citenamefont {Brown~Jr.}(1963)}]{brown}%
  \BibitemOpen
  \bibfield  {author} {\bibinfo {author} {\bibfnamefont {W.~F.}\ \bibnamefont {Brown~Jr.}},\ }\href@noop {} {\emph {\bibinfo {title} {Micromagnetics}}}\ (\bibinfo  {publisher} {Interscience Publishers},\ \bibinfo {address} {New York},\ \bibinfo {year} {1963})\BibitemShut {NoStop}%
\bibitem [{\citenamefont {Aharoni}(2000)}]{aharonibook}%
  \BibitemOpen
  \bibfield  {author} {\bibinfo {author} {\bibfnamefont {A.}~\bibnamefont {Aharoni}},\ }\href@noop {} {\emph {\bibinfo {title} {{Introduction to the Theory of Ferromagnetism}}}},\ \bibinfo {edition} {2nd}\ ed.\ (\bibinfo  {publisher} {Oxford University Press},\ \bibinfo {address} {Oxford},\ \bibinfo {year} {2000})\BibitemShut {NoStop}%
\bibitem [{Note1()}]{Note1}%
  \BibitemOpen
  \bibinfo {note} {In the approach-to-saturation regime, $d\Sigma _{\protect \mathrm {M}} / d \Omega $ equals the total (nuclear and magnetic) unpolarized SANS cross section minus the total SANS cross section at complete magnetic saturation.}\BibitemShut {Stop}%
\bibitem [{\citenamefont {Michels}\ \emph {et~al.}(2003)\citenamefont {Michels}, \citenamefont {Viswanath}, \citenamefont {Barker}, \citenamefont {Birringer},\ and\ \citenamefont {Weissm\"uller}}]{michels03prl}%
  \BibitemOpen
  \bibfield  {author} {\bibinfo {author} {\bibfnamefont {A.}~\bibnamefont {Michels}}, \bibinfo {author} {\bibfnamefont {R.~N.}\ \bibnamefont {Viswanath}}, \bibinfo {author} {\bibfnamefont {J.~G.}\ \bibnamefont {Barker}}, \bibinfo {author} {\bibfnamefont {R.}~\bibnamefont {Birringer}},\ and\ \bibinfo {author} {\bibfnamefont {J.}~\bibnamefont {Weissm\"uller}},\ }\bibfield  {title} {\bibinfo {title} {{Range of Magnetic Correlations in Nanocrystalline Soft Magnets}},\ }\href {https://doi.org/10.1103/PhysRevLett.91.267204} {\bibfield  {journal} {\bibinfo  {journal} {Phys. Rev. Lett.}\ }\textbf {\bibinfo {volume} {91}},\ \bibinfo {pages} {267204} (\bibinfo {year} {2003})}\BibitemShut {NoStop}%
\bibitem [{\citenamefont {Bick}\ \emph {et~al.}(2013{\natexlab{a}})\citenamefont {Bick}, \citenamefont {Honecker}, \citenamefont {D\"obrich}, \citenamefont {Suzuki}, \citenamefont {Gilbert}, \citenamefont {Frielinghaus}, \citenamefont {Kohlbrecher}, \citenamefont {Gavilano}, \citenamefont {Forgan}, \citenamefont {Schweins}, \citenamefont {Lindner}, \citenamefont {Birringer},\ and\ \citenamefont {Michels}}]{bickapl2013}%
  \BibitemOpen
  \bibfield  {author} {\bibinfo {author} {\bibfnamefont {J.-P.}\ \bibnamefont {Bick}}, \bibinfo {author} {\bibfnamefont {D.}~\bibnamefont {Honecker}}, \bibinfo {author} {\bibfnamefont {F.}~\bibnamefont {D\"obrich}}, \bibinfo {author} {\bibfnamefont {K.}~\bibnamefont {Suzuki}}, \bibinfo {author} {\bibfnamefont {E.~P.}\ \bibnamefont {Gilbert}}, \bibinfo {author} {\bibfnamefont {H.}~\bibnamefont {Frielinghaus}}, \bibinfo {author} {\bibfnamefont {J.}~\bibnamefont {Kohlbrecher}}, \bibinfo {author} {\bibfnamefont {J.}~\bibnamefont {Gavilano}}, \bibinfo {author} {\bibfnamefont {E.~M.}\ \bibnamefont {Forgan}}, \bibinfo {author} {\bibfnamefont {R.}~\bibnamefont {Schweins}}, \bibinfo {author} {\bibfnamefont {P.}~\bibnamefont {Lindner}}, \bibinfo {author} {\bibfnamefont {R.}~\bibnamefont {Birringer}},\ and\ \bibinfo {author} {\bibfnamefont {A.}~\bibnamefont {Michels}},\ }\bibfield  {title} {\bibinfo {title} {{Magnetization reversal in Nd-Fe-B based nanocomposites as seen by magnetic small-angle neutron scattering}},\
  }\href {https://doi.org/10.1063/1.4776708} {\bibfield  {journal} {\bibinfo  {journal} {Appl. Phys. Lett.}\ }\textbf {\bibinfo {volume} {102}},\ \bibinfo {pages} {022415} (\bibinfo {year} {2013}{\natexlab{a}})}\BibitemShut {NoStop}%
\bibitem [{\citenamefont {Weissm\"uller}\ \emph {et~al.}(2001)\citenamefont {Weissm\"uller}, \citenamefont {Michels}, \citenamefont {Barker}, \citenamefont {Wiedenmann}, \citenamefont {Erb},\ and\ \citenamefont {Shull}}]{jprb2001}%
  \BibitemOpen
  \bibfield  {author} {\bibinfo {author} {\bibfnamefont {J.}~\bibnamefont {Weissm\"uller}}, \bibinfo {author} {\bibfnamefont {A.}~\bibnamefont {Michels}}, \bibinfo {author} {\bibfnamefont {J.~G.}\ \bibnamefont {Barker}}, \bibinfo {author} {\bibfnamefont {A.}~\bibnamefont {Wiedenmann}}, \bibinfo {author} {\bibfnamefont {U.}~\bibnamefont {Erb}},\ and\ \bibinfo {author} {\bibfnamefont {R.~D.}\ \bibnamefont {Shull}},\ }\bibfield  {title} {\bibinfo {title} {{Analysis of the small-angle neutron scattering of nanocrystalline ferromagnets using a micromagnetics model}},\ }\href {https://doi.org/10.1103/PhysRevB.63.214414} {\bibfield  {journal} {\bibinfo  {journal} {Phys. Rev. B}\ }\textbf {\bibinfo {volume} {63}},\ \bibinfo {pages} {214414} (\bibinfo {year} {2001})}\BibitemShut {NoStop}%
\bibitem [{\citenamefont {Bick}\ \emph {et~al.}(2013{\natexlab{b}})\citenamefont {Bick}, \citenamefont {Suzuki}, \citenamefont {Gilbert}, \citenamefont {Forgan}, \citenamefont {Schweins}, \citenamefont {Lindner}, \citenamefont {K\"ubel},\ and\ \citenamefont {Michels}}]{bickapl2013no2}%
  \BibitemOpen
  \bibfield  {author} {\bibinfo {author} {\bibfnamefont {J.-P.}\ \bibnamefont {Bick}}, \bibinfo {author} {\bibfnamefont {K.}~\bibnamefont {Suzuki}}, \bibinfo {author} {\bibfnamefont {E.~P.}\ \bibnamefont {Gilbert}}, \bibinfo {author} {\bibfnamefont {E.~M.}\ \bibnamefont {Forgan}}, \bibinfo {author} {\bibfnamefont {R.}~\bibnamefont {Schweins}}, \bibinfo {author} {\bibfnamefont {P.}~\bibnamefont {Lindner}}, \bibinfo {author} {\bibfnamefont {C.}~\bibnamefont {K\"ubel}},\ and\ \bibinfo {author} {\bibfnamefont {A.}~\bibnamefont {Michels}},\ }\bibfield  {title} {\bibinfo {title} {{Exchange-stiffness constant of a Nd-Fe-B based nanocomposite determined by magnetic neutron scattering}},\ }\href {https://doi.org/10.1063/1.4821453} {\bibfield  {journal} {\bibinfo  {journal} {Appl. Phys. Lett.}\ }\textbf {\bibinfo {volume} {103}},\ \bibinfo {pages} {122402} (\bibinfo {year} {2013}{\natexlab{b}})}\BibitemShut {NoStop}%
\bibitem [{Note2()}]{Note2}%
  \BibitemOpen
  \bibinfo {note} {Note that asymptotically $R_{\protect \mathrm {H}} \propto q^{-4}$ and $R_{\protect \mathrm {M}} \propto q^{-2}$ in Eq.~(\ref {sigmasmperp}), which together with the (unknown) $q$~dependencies of $S_{\protect \mathrm {H}}$ and $S_{\protect \mathrm {M}}$ result in $d \Sigma _{\protect \mathrm {M}} / d \Omega \propto q^{-n}$ with $n > 4$.}\BibitemShut {Stop}%
\end{thebibliography}

%

\end{document}